\documentclass[conference]{IEEEtran}
\usepackage{hyperref}
\usepackage[table,xcdraw]{xcolor}
\usepackage{balance}
\usepackage{cite}
\ifCLASSINFOpdf
   \usepackage[pdftex]{graphicx}
\else
\fi
\begin{document}
\title{Critical Requirements Engineering in Practice}
% maybe the "IN" is not so bad

\author{\IEEEauthorblockN{Leticia Duboc}
\IEEEauthorblockA{Department of Engineering\\
 La Salle. Ramon Llull University\\
 Spain\\
Email:  l.duboc@salle.url.edu}
\and
\IEEEauthorblockN{Curtis McCord, Christoph Becker}
\IEEEauthorblockA{ Faculty of Information\\
University of Toronto \\
Canada\\
Email:\{curtis.mccord, christoph.becker\}@utoronto.ca}
\and
% \IEEEauthorblockN{Christoph Becker}
% \IEEEauthorblockA{ Faculty of Information\\
% University of Toronto \\
% Canada\\
% Email:christoph.becker@utoronto.ca }
\and
\IEEEauthorblockN{Syed Ishtiaque Ahmed}
\IEEEauthorblockA{Department of Computer Science\\
University of Toronto \\
Canada\\
Email: ishtiaque@cs.toronto.edu }
}

\maketitle

%\begin{abstract}
%Software have a great potential to for improving human lives. It can also enable all sorts of maladies, raising issues of ethics, power, politics, and human and social values. The bet opportunity to deal with such issues during software development is within Requirements Engineering (RE). Yet, many RE frameworks lack the concepts to address lack the concepts to address politics, morality, aesthetics, and beliefs. In this paper, we should how Critical System Heuristics (CSH) can be used for structuring early explorations of requirements, providing a framework for developing a reflective understanding for justifying the project and system scope. 
%\end{abstract}

%requirements engineering, critical system heuristics, system thinking 

% \fi
\IEEEpeerreviewmaketitle

\section{Introduction}
Today's societies run on software. Software Engineering (SE) is rightfully expected to address social and ethical concerns of software systems in society. Software systems have enormous potential for human improvement, but are implicated in all sorts of maladies. From radicalizing voters to the erosion of privacy, from mechanisms to support emission test cheating to the energy impact of bitcoin mining, software is never a neutral element of society \cite{eubanks} \cite{safiyanoble}. This raises serious ethical issues, including  responsibility and power relationships in systems design, the politics of stakeholder engagement, and the role of human and social values in engineering \cite{feenberg_ten_2010}. 
Within SE, Requirements Engineering (RE) is arguably the key leverage point for social change and sustainability \cite{becker_requirements:_2016}. By focusing on whom to involve as stakeholders, how to elicit their perspectives, how to consider these in architectural design choices, and how to define acceptance criteria, RE frames the design scope and establishes the success conditions for systems development. It is no surprise then that it is increasingly called upon to address social and ethical concerns of software systems in society \cite{ruhe_vision:_2017}. In practice, RE and SE professionals must engage a broad range of stakeholders, facilitate the emergence of a shared understanding of what the systems development effort should achieve, and represent the outcomes of that understanding. In this process, they carry an ethical and moral responsibility to all those affected, and in particular, to those affected but not involved in systems design. This means that they must engage with the normative frameworks and rules of \textit{ethics}; pay attention to how human \textit{values} -- what people regard as important -- constitute the `facts of the future'\cite{feenberg_ten_2010}; understand how their own work is subject to social relationships of stakeholders (\textit{politics}); and be sensitive to issues of \textit{power}: who it is held by, how it is wielded and in which form it influences choices and technological trajectories. This is challenging, because software systems are fully intertwined with physical and natural environments, economic processes, and with social and cultural life \cite{jarke_brave_2011}. However, as many Requirements Engineering frameworks are ultimately grounded in the scientific method, they lack the concepts to address politics, morality, aesthetics, and beliefs \cite{churchman_systems_1979,ulrich1983critical}. This makes it so difficult to move between what Goguen called the ``wet'' and the ``dry'' -- between rich human and social worlds and the formalized technical models and methods used in software engineering \cite{goguen_requirements_1994}. 

To address the social construction of requirements, RE often suggests the use of interpretive systems thinking frameworks such as Soft Systems Methodology, which focuses on facilitating the emergence of shared understanding \cite{checkland_systems_1999}. But these frameworks are not able to address the marginalization that inevitably arises out of power dynamics \cite{jackson2007systems}. The need to make visible and reflect on such concerns \cite{ulrich1983critical} led to the development of Critical Systems Thinking (CST) frameworks, but only a few have considered them in RE \cite{wing2015systemic} \cite{elsawah2015beyond}. In part, this can be attributed to their focus on philosophical theory, social critique, and epistemology. 

As a result, practitioners can feel rather helpless. Even if they have the best intentions, how are they to ``rationally justify the normative implications of systems design'', as Ulrich framed it \cite{ulrich1983critical}? In other words, how can they justify their work, their design decisions, their actions, when it is not considered feasible to estimate or predict possible effects spread in time and space; when they have no foundational education in social sciences, policy, or ethics; when they are embedded in industry projects with tight time lines, profit expectations and dispersed networks of potential stakeholders?

\begin{figure*}
\centering
\includegraphics[width=16cm]{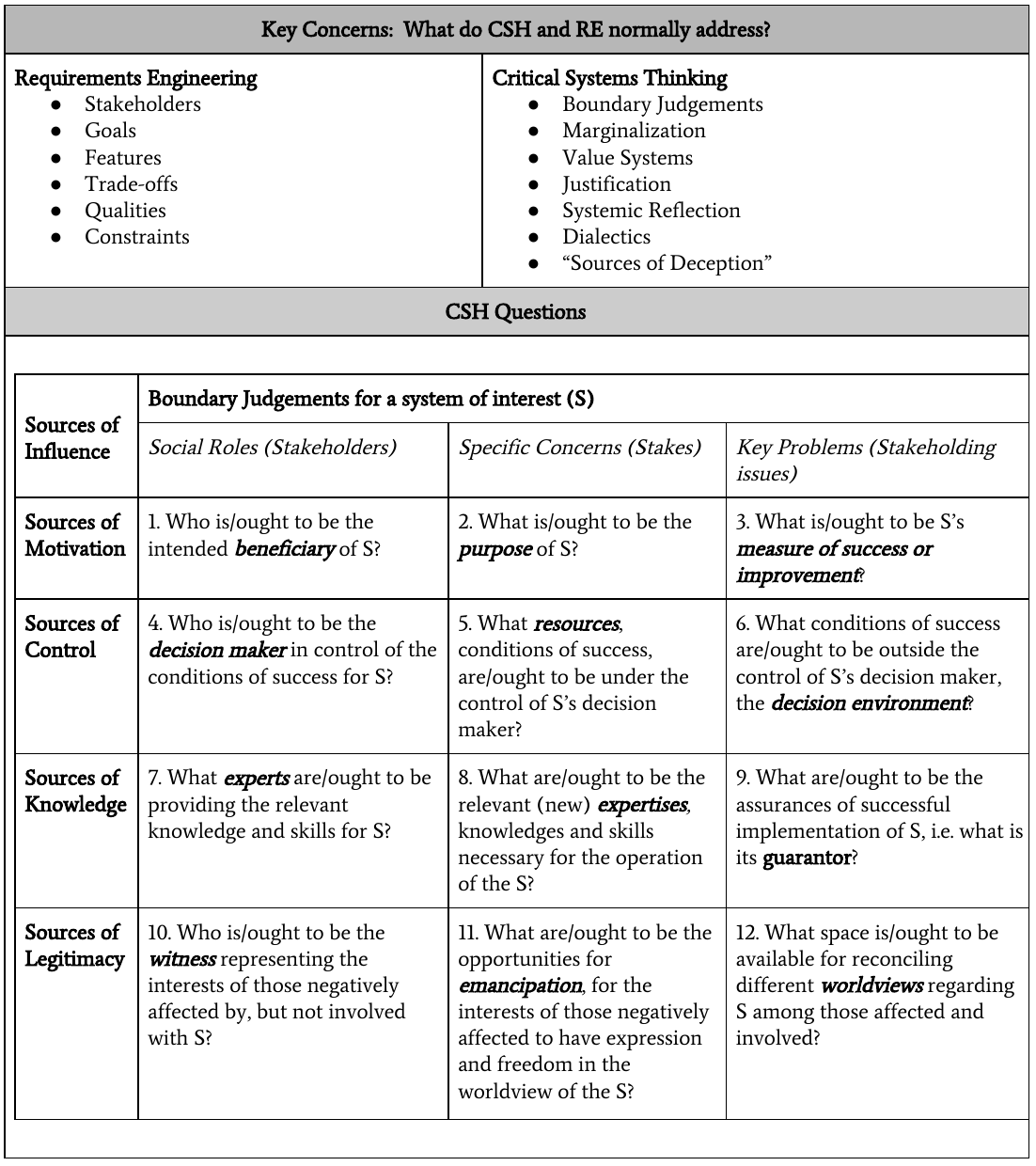}
% where an .eps filename suf fix will be assumed under latex, 
% and a .pdf suffix will be assumed for pdflatex; or what has been declared
% via \DeclareGraphicsExtensions.
\caption{Overview of CSH  and Key Concerns, adapted from \cite{ulrich2010critical}.}
%TODOS: clarify; expand! also, can we balance and better cross-map? e.g. goals <> value systems; (add) prioritization <> marginalization; trade-offs <> dialectics
\label{CSH Questions}
\end{figure*}

In this article, we describe a collaborative \textit{Action Research} project\cite{checkland_action_1998} that demonstrates how Critical Systems Heuristics (CSH), the best known CST framework (see sidebar \ref{cshsb}), can be used in the context of RE to gain a critical awareness of power and politics; to support critical reflection on the human values and boundary judgments that guide and frame a project; and to support requirements professionals in understanding and questioning the selectivity of their choices. Intertwining CSH with standard RE practice helped us to make visible the risk of marginalization faced by the beneficiaries in a software-driven technology development project, and to adjust early requirements activities to more fairly represent the concerns of those at risk of marginalization. As a result, the project created two types of artifacts: A requirements document following the often-used Volere template \cite{VolereTemplate}, and a set of what CSH calls `\textit{ideal maps}'-- accounts of the values, knowledge, politics and perspectives forming the basis of the requirements statements. Fig. \ref{CSH Questions} summarizes the kinds of issues normally addressed by these two frameworks as well as the CSH questions guiding the creation of these maps. 
Our findings show that CSH is invaluable in supporting RE. CSH lacks the content and structure of RE, while RE practice lacks an approach to critical reflection on the values driving systems development. When combined effectively, the result is a critically appreciative RE practice that can be adopted by software professionals.

\section{Combining RE and CSH in software system development}

\subsection{The problem situation}
\noindent The world's population is aging. In Spain, where this project takes place, 34,6\% of the population is expected to be above 65 years by 2066. Many will live alone and may require support. Clues that they need help can be subtle, and may pass unnoticed by their families and caretakers. For example, a change in routine could indicate the beginning of certain diseases like dementia\cite{hayes2008unobtrusive}. These clues would either have to be observed by visitors, who are only present for a small portion of the day, or communicated by the elderly people themselves, who may not have the cognitive capacity to notice the cues or realize when they indicate a problem. 

\subsection{The Project}
The HomeSound project focuses on a vulnerable population of fragile elderly people living at home. It aims to design a Wireless Acoustic Sensor Network and algorithms to support their lives by detecting unusual sounds that might indicate possible changes in routine, accidents, or mishaps (see sidebar \ref{aedsb}). The project is in the early stages of RE, focused on exploring the potential for technology transfer.

\begin{table*}
%\begin{longtable*}[!t]
%\normalsize
\caption{Iterations of the ideal map}
\label{table}
\centering
\begin{tabular*}{\textwidth}{|c||p{16cm}|}
%\begin{supertabular*}{\textwidth}{|c||p{16.05cm}|}
\hline
Ideal Map & Description\\
\hline
1 & Ideal map 1 was created by the requirements engineer and one of the CSH experts, as the latter walked the former through the CSH process. It contained the requirements engineer's observations from their previous discussions with the HomeSound development team, third-sector organizations, and the families and caretakers of elderly people. Among other things, the map identified the elderly as the main \textit{beneficiary}, and the stated \textit{purpose} as increasing their independence, to allow them to stay home longer. It also set the elderly's `independence and well-being levels' and the `number of years living alone at home' as \textit{measures of improvement}. We also assumed that if their well-being increased, this would reduce the  `number of distressed calls from elderly people' to their caretakers, which could therefore be taken as a possible \textit{guarantor of success}. %Space constraints prevents us from discussing the maps in detail. 

Reflection revealed, among other things, that this first map was built from the \textbf{selective memory} of the requirement engineer and most likely represented the \textbf{privileged views} of those developing the system. We therefore decided to consult the notes of previous conversations with stakeholders to more accurately reflect their needs. We also felt the need to involve a third researcher on the critique of the maps, and to help reconcile and synthesize the dialogue in lieu of more thorough participant involvement. \\
\hline

2 & Ideal map 2 was created after the consultation of the notes from previous meetings and interviews with stakeholders. The new map added the families as a \textit{primary beneficiary} and the society and the healthcare system as \textit{secondary beneficiaries}, extending the \textbf{boundaries} of the system. The \textit{purpose} now  included peace of mind for families. Ideal map 2 also reviewed the concept of well-being and independence, among other things.

Reflection  on the validity of our \textit{measurements of success} made us realize that we needed to get input from professionals with \textit{expertise} relevant to the situations of elderly people. For Ulrich, relying on incomplete or dogmatic perspectives is a major `\textbf{source of deception}' and can be a \textit{false guarantor} that harms our understanding of the situation and our systems design \cite{ulrich1983critical}. Reflection also raised a number of questions, including: What are the boundaries of the system? Is `years at home' a suitable measure of well-being? What about uninvolved family members? Should the purpose be to increase independence or rather self-determination? Do people really understand the implications of the technology? Why have we observed a blind trust in technology in our conversations with stakeholders? What if the elderly cannot be the decision makers? \textit{etc}. This last question in particular raises important issues of \textit{fairness in  representing the concerns of those at risk of marginalization}.\\
\hline
3 & Ideal map 3 integrated the views of a social worker and a psychologist, both specialized in the isues of the elderly. The interviews highlighted several issues, including that the system did not increase independence,  but rather security (as it cannot meet their physical and emotional needs) and that the number of calls from the elderly were a \textbf{false guarantor of success}. We also learned about common behaviors, coping mechanisms, the importance of a trusted person, and several scales for measuring well-being of elderly people and their caretakers. Ideal map 3 included new \textit{ measures of success} (e.g. social support, anxiety of the caretaker, early signs of dementia) and professional scales used in psychology and social care to measure them, and an extended list of\textit{ decision makers} and \textit{sources of knowledge}.

Reflection raised doubts regarding the measurement of self-determination and early signs of dementia. %, and brought up questions such as: Is RE missing critical reflection? What should the the role of  CSH in RE be?
\\
\hline
4 & Ideal map 4  was developed after an interview with a practical philosopher specialised on the impact of technological projects on ethics and privacy. We discussed issues such as: How do you frame care and well-being? What is the amount of trust required from the user? Can over-reliance of families on this technology lead to a loss of `human touch' and thus reduce well-being rather than support it? Can the technology reduce the autonomy of the elderly, who should have the right to decide when to get help? Will third parties be interested in this data? \textit{etc}. Each of these questions bring \textbf{power} imbalance and the \textbf{politics} of stakeholders to the forefront of RE activities. Finally, we recognized the importance of public debate on such technologies, and we identified techniques from practical philosophy for uncovering stakeholder ethics, morals and values.  The new map included  autonomy as a \textit{primary aim}, a better definition of self-determination, the general public as a \textit{desired expert}, institutions that are interested in data as a commodity as an \textit{undesired expert}, and possible \textit{worldviews} about being old, supporting the elderly, living the good life, and surveillance technology.

Reflection led us to recognize we had been more concerned with people's perception than with security, assuming this was something more easily solvable. Hence, we chose to interview a security expert.
\\
\hline
5 & Ideal map 5 integrated the opinions of an IT security specialist who performs security audits of technological solutions. They highlighted that sound monitoring may not be the best solution to the problem due to the intrinsic risks of having microphones in the house. We learned about the different privacy and security risks of these devices, and were recommended a security audit when the system is under development. The new map included the \textit{desired skills} that a security expert could bring to the project, a new \textit{guarantor of success} (the number of vulnerabilities discovered in recurrent security audits), and several risks and possible measures.

Reflection led us to a discussion on the role of risk assessment in CSH and whether it could lead to overlooking the important issue of moral justification for the project. We also reflected on how the insights from CSH were different from our previous experiences with RE techniques, and on how we could integrate these insights into RE. This led us to attempt to map the CSH findings to the Volere framework. 
\\
\hline
6 & Ideal map 6 incorporated the insights derived from translating the findings of the CSH to the Volere specification. These included classifying previous secondary aims as \textit{unrealistic aims} (as the technical system could only address specific aspects of this larger aim) and improving rationale, relationship, and consistency for items across different sections of the ideal map. 

Reflection showed us that there is value in iterating between standard RE and CSH, as discussed in Section \ref{discussion}.  %Creating the CSH ideal map and the Volere specification enabled us to incorporate the critical reflection that RE frameworks on their own do not provide. In order to facilitate the integration of CSH and RE, we  created an ideal map template complementing Ulrich’s questions, a mapping to from this template to the Volere template,  and an annotated version of the latter. \footnote{The first two are available in \underline{http://tiny.cc/cnbj7y} and  \underline{http://tiny.cc/okbj7y}. Documents will be moved to ArXiv if the paper is accepted. The latter cannot be made available due to copyright issues.}
   \\
\hline
\end{tabular*}
%\end{supertabular*}
%\end{longtable*}
\end{table*}

\subsection{The Process}
In this Action Research, we have combined standard RE practice with CSH through iterative cycles of critical RE practice followed by reflection. The research team was composed of a researcher at La Salle and two from the University of Toronto. The former had been involved in conversations with third-sector companies and other stakeholders and has close contact with the technical and business developers of HomeSound, but had no previous knowledge of CSH. The other two were knowledgeable in CSH. They took a mentoring role and helped to critically reflect on the models created by the first researcher. We will refer to these as requirements engineer and CSH experts, respectively. To make the activities and findings \textit{recoverable}\cite{checkland_action_1998} within the space constraints of the article, we provide a high-level overview of the iterations and have made the template materials available (see below).

The requirements engineer created several versions of the ideal map for the HomeSound project, guided by the questions of Table \ref{table}. As in any iterative process, ideal maps are incomplete. The first two maps, for example, reflect the privileged views of those building the systems. These views are slowly extended to incorporate the viewpoint of different stakeholders and the results of the team's critical reflection on that version of the map. Table \ref{table} summarizes this process, highlighting in italics the topics that refer to the CSH questions (see Fig. \ref{CSH Questions}), and in bold the kind of awareness commonly brought up  by CSH\footnote{The CSH process still continues, now to integrate the views of elderly people. But at the time of writing no new version of the ideal map had been generated from the later inputs.}.

%\subsection{Bringing CSH and RE together}

%Creating the CSH ideal map and the Volere specification enabled us to incorporate the critical reflection that RE frameworks on their own do not provide. In order to facilitate the integration of CSH and RE, we  created an ideal map template complementing Ulrich’s questions, a mapping to from this template to the Volere template,  and an annotated version of the latter. \footnote{The first two are available in \underline{http://tiny.cc/cnbj7y} and  \underline{http://tiny.cc/okbj7y}. Documents will be moved to ArXiv if the paper is accepted. The latter cannot be made available due to copyright issues.  }

\section{Discussion and Conclusion}
\label{discussion}
%Same sentence used in the introduction:
%Software Engineering is increasingly called on to address social and ethical concerns of software systems in society. This raises issues of power and politics and human and social values that must primarily be addressed through Requirements Engineering activities. 
Requirements engineering must address issues of power, politics and human social values. Critical Systems Thinking frameworks have been developed to make visible and reflect on such concerns, but have not been taken up in RE. 
This article asked:  \textit{What is the role of Critical Systems Heuristics in Requirements Engineering? }

The process described in Table \ref{table} illustrates the challenges that the \textit{engineering} perspectives of SE face. The systems that matter when software is developed are inevitably socio-technical systems. Abdicating the responsibility to account for that is not an ethical option, but it is difficult to do justice to the implications that result. As others have argued, RE is in a unique position to address the social responsibility that derives from software systems' central role in society \cite{becker_requirements:_2016,ruhe_vision:_2017}. Living up to that role requires integrating social theory and critical perspectives into the core of what RE does, and how it accounts for what it does.

In this article, we showed how Critical Systems Heuristics can be used for structuring early explorations of requirements in a project designing an embedded device that supports elderly people at home. We proceeded in iterative progressions through consecutive versions of a map of critical categories, using interviews, reflection, and internal critique. The role of the CSH framework was to provide an effective framing for developing a reflexive understanding of stakeholders and concerns; revising high-level purposes, goals and measures of success to represent the interests of those affected; probing from multiple perspectives how different project and system scopes can and should be justified; and exploring how human, social and economic values should drive the project. Some of the issues raised during the process were highlighted in bold and italics in Table \ref{table}. For example, we learned that, if not carefully designed, the system could reduce the autonomy of the elderly-- the very opposite of our intended purpose. We also revisited our initial understanding of well-being and elderly support, leading us to balance ``safety'' and ``self-determination''. These issues had not been challenged by the privileged view of the system's developers prior to this exercise. 

This type of critical reflection complements recent work on values and politics in RE that explicitly considered human values in the RE process \cite{thew2018value}, and modelled stakeholder interactions \cite{milne2012power} in a way similar to the political analyses of Soft Systems Methodology \cite{checkland_systems_1999}. Rather than focus on a descriptive analysis of power and politics in systems design, as do Milne and Maiden \cite{milne2012power}, CST and CSH commit practitioners to revealing, and even avoiding, situations where power can marginalize perspectives of presumed beneficiaries. Following this, we focus on discursive acts, and supporting critical reflection that makes visible the implicit boundary judgments of all involved. In attitude, this is closer to critical design \cite{dunne2013speculative} and similar approaches in HCI. But because CSH, despite its specific language and terminology, is a small-scale heuristic framework, it is highly suitable for being adopted quickly and without extensive study of social theory. 

%such as Situated Intervention  \cite{zuiderent-jerak_situated_2015}

Beyond a requirements specification, using CSH in the project also created \textit{ideal maps}. Translating the issues captured by the ideal maps to the Volere specification forced us to think in more concrete terms about the system. We realized, for example, that the project's secondary purposes of ``Better serving the society by allowing elderly people to stay longer at their own home'' and ``Reducing the financial burden of the social security system'' really fall outside of a reasonable and justifiable design scope, being in fact an \textit{unrealistic aim}, though the technical contribution can and should address specific aspects of this larger aim.

Although we can easily map the CSH questions to the Volere template, the type of issues that the process revealed are of a very different nature from those typically found in requirements documents, including educational sample specifications such as the Volere package. The latter are more focused on the system features than the values of the stakeholders involved and affected. While we are not claiming that the analysis of real system is a fair comparison to a sample specification, it does makes us wonder if the RE community should deliberately create didactic materials that highlight issues of ethics, power, politics, and human and social values.

Finally, creating the CSH ideal map and the Volere specification enabled us to incorporate the critical reflection that RE frameworks on their own do not provide. In order to facilitate the integration of CSH and RE, we  created an ideal map template complementing Ulrich´s questions, a mapping from this template to the Volere template,  and an annotated version of the latter.\footnote{Refer to: \url{https://www.sustainabilitydesign.org/publications/\#materials}. The latter cannot be made available due to copyright issues.}

RE is a key leverage point for addressing social and ethical concerns of software systems in society. We have demonstrated the value of Critical Systems Heuristics in early-phase RE. While this cannot guarantee ethical and fair software systems design - nothing can - it provides a crucial stepping stone for the different kind of SE that is called for in the 21st century. 

\begin{table*}[!h]
\caption{SIDE BAR: Critical System Heuristics. }
\label{cshsb}
\centering
\begin{tabular*}{\textwidth}{|p{18cm}|}
\hline

% ATTENTION: THIS IS THE VERSION CURTIS HAD IN THE SPREADSHEET, HOWEVER, I DON'T KNOW IF THE TEXT THAT IS NOW COMMENTED HAD BEEN EDITED BY CHRISTOPH (I BELIEVE IT IS NOT). 

Critical Systems Heuristics (CSH) is a methodology and approach to decision-making and evaluation within systems design. CSH emphasizes that the boundaries constituting a system are not objective necessities, but enacted through decisions. Claims about scope, measurement, or stakeholders are normative, value-laden claims that must be not (just) optimal, but also legitimate. At stake in any situation are the conditions by which these decisions are made. 

Beliefs in the objectivity of expertise or the exigencies of powerful stakeholders can lead to the marginalization of lay voices in decision-making. CSH is an ethical commitment to participatory design, and to creating conditions where those affected by a system can be involved in planning it. When experts and lay-participants interact, there are asymmetries in knowledge and power, but the burden of explanation lies with the experts involved. The situated knowledge and values of those affected are in-themselves qualification to speak freely and critically about the assumptions and judgments of those with power in decision making and design. 

CSH focuses on revealing the values underlying design by iteratively applying 12 questions (see Fig. \ref{CSH Questions}) covering dimensions of \textbf{motivation}, \textbf{control}, \textbf{knowledge}, and \textbf{legitimization}. Iterations alternate between descriptive (`is') and ideal (`ought to be') modes. The goals of CSH are to make the values shaping the scope of the system visible; to allow those involved in design to reflect on beliefs supporting purpose, practise and improvement; and to create a space where  those affected by design and implementation are on equal footing with the knowledge of experts.

\\
\hline
\end{tabular*}
\end{table*}

\begin{table*}[!h]
\caption{SIDE BAR: Acoustic Event Detection.}
\label{aedsb}
\centering
\begin{tabular*}{\textwidth}{|p{18cm}|}
\hline
\\
\textbf{Acoustic Event Detection} automatically identifies events of interest from within continuous audio streams \cite{temko2007acoustic}. This technology has several applications, like home security (e.g.\ CO2 alarm, broken window), people and pet care (e.g.\ routine change, barking), transportation (e.g.\ traffic monitoring), entertainment (e.g.\ adapt sound spectrum based on surrounding noise), wellness (e.g.\ baby crying, snoring), social (e.g.\ pause music when having conversation).
% \cite{jeronimo2017driving}

The \textit{SmartSound} \cite{socoro2017anomalous} is an Anomalous Noise Event (ANE) Detector designed to work in real-time on low-cost acoustic sensors of a Wireless Acoustic Sensor Network (WASN). It uses Gaussian Mixture Modelling to distinguish anomalous noise events from background noise. Its most mature application is in the real-time detection and representation of the acoustic impact of road infrastructure. It has also been used to monitor endangered bird species in a national park, and on an Ambient Assisted Living platform to assist medical staff to track the status of patients in real-time.
\\
\hline
\end{tabular*}
\end{table*}

%\begin{table*}[!t]
%\caption{SIDE BAR: Related Work}
%\centering
%\begin{tabular*}{\textwidth}{|p{18cm}|}
%\hline
% Human-computer Interaction (HCI) has long studied technologies and processes to better conceptualize and accommodate various political tensions surrounding a computing systems. Within HCI, critical thinking has made its way at least through two different veins: (a) Critical Design, and (b) Critical Technical practices. Inspired from Critical Art, Critical design proposes a reflective framework that enables the users to reflect on their use. Critical design is hence a tool that defamiliarize an object or a phenomenon to a person. This tool is particularly important and is often used in the early phase of design when various stakeholders contend on an issue and a persuasion requires them to perceive a situation from a different perspective. On the other hand, Critical Technical Practice shifts the focus from design to the social context in which the designed artifact will be used. This framework emphasizes on the fact that, despite the intention of the designers, the use of a technology is shaped by the social and cultural context, and thus creates a tension between the designed and perceived values. Taken together, the critical discourse within HCI creates a dialectic tension around the material practices of design, and brings to the fore the power dynamics of the designer and the stakeholders. \\
%\hline
%\hline
%\end{tabular*}
%\end{table*}

%\FloatBarrier

 \section*{Acknowledgment}
 The research leading to these results has received funding from the European Union’s Horizon 2020 research and innovation programme under the Marie Skłodowska-Curie grant agreement No 712949 (TECNIOspringPLUS), from the Agency for Business Competitiveness of the Government of Catalonia, and from NSERC under RGPIN-2016-06640.
 
\balance
 
\bibliographystyle{IEEEtran}
\bibliography{homesound}

\begin{thebibliography}{10}
\providecommand{\url}[1]{#1}
\csname url@samestyle\endcsname
\providecommand{\newblock}{\relax}
\providecommand{\bibinfo}[2]{#2}
\providecommand{\BIBentrySTDinterwordspacing}{\spaceskip=0pt\relax}
\providecommand{\BIBentryALTinterwordstretchfactor}{4}
\providecommand{\BIBentryALTinterwordspacing}{\spaceskip=\fontdimen2\font plus
\BIBentryALTinterwordstretchfactor\fontdimen3\font minus
  \fontdimen4\font\relax}
\providecommand{\BIBforeignlanguage}[2]{{%
\expandafter\ifx\csname l@#1\endcsname\relax
\typeout{** WARNING: IEEEtran.bst: No hyphenation pattern has been}%
\typeout{** loaded for the language `#1'. Using the pattern for}%
\typeout{** the default language instead.}%
\else
\language=\csname l@#1\endcsname
\fi
#2}}
\providecommand{\BIBdecl}{\relax}
\BIBdecl

\bibitem{eubanks}
V.~Eubanks, \emph{Automating Inequality}.\hskip 1em plus 0.5em minus
  0.4em\relax St. Martin's Press, 2018.

\bibitem{safiyanoble}
S.~U. Noble, \emph{Algorithms of Oppression}.\hskip 1em plus 0.5em minus
  0.4em\relax NYU Press, 2018.

\bibitem{feenberg_ten_2010}
A.~Feenberg, ``Ten paradoxes of technology,'' \emph{Techné: Research in
  Philosophy and Technology}, vol.~14, no.~1, pp. 3--15, 2010.

\bibitem{becker_requirements:_2016}
C.~Becker, S.~Betz, R.~Chitchyan, L.~Duboc, S.~M. Easterbrook,
  B.~Penzenstadler, N.~Seyff, and C.~C. Venters, ``Requirements: {The} {Key} to
  {Sustainability},'' \emph{IEEE Software}, vol.~33, no.~1, pp. 56--65, Jan.
  2016.

\bibitem{ruhe_vision:_2017}
G.~Ruhe, M.~Nayebi, and C.~Ebert, ``The {Vision}: {Requirements} {Engineering}
  in {Society},'' in \emph{2017 {IEEE} 25th {International} {Requirements}
  {Engineering} {Conference} ({RE})}, Sep. 2017, pp. 478--479.

\bibitem{jarke_brave_2011}
\BIBentryALTinterwordspacing
M.~Jarke, P.~Loucopoulos, K.~Lyytinen, J.~Mylopoulos, and W.~Robinson, ``The
  {Brave} {New} {World} of {Design} {Requirements},'' \emph{Information
  Systems}, vol.~36, no.~7, pp. 992--1008, Nov. 2011. [Online]. Available:
  \url{http://dx.doi.org/10.1016/j.is.2011.04.003}
\BIBentrySTDinterwordspacing

\bibitem{churchman_systems_1979}
C.~W. Churchman, \emph{\BIBforeignlanguage{en}{The systems approach and its
  enemies}}.\hskip 1em plus 0.5em minus 0.4em\relax Basic Books, May 1979.

\bibitem{ulrich1983critical}
W.~Ulrich, \emph{Critical Heuristics of Social Planning: A New Approach to
  Practical Philosophy}.\hskip 1em plus 0.5em minus 0.4em\relax J. Wiley and
  Sons, 1983.

\bibitem{goguen_requirements_1994}
J.~Goguen, ``Requirements {Engineering} as the {Reconciliation} of {Technical}
  and {Social} {Issues},'' in \emph{Requirements {Engineering}: {Social} and
  {Technical} {Issues}}.\hskip 1em plus 0.5em minus 0.4em\relax Academic Press,
  1994, pp. 165--199.

\bibitem{checkland_systems_1999}
P.~Checkland, \emph{\BIBforeignlanguage{English}{Systems {Thinking}, {Systems}
  {Practice}: {Includes} a 30-{Year} {Retrospective}}}.\hskip 1em plus 0.5em
  minus 0.4em\relax Wiley, 1999.

\bibitem{jackson2007systems}
M.~C. Jackson, \emph{Systems approaches to management}.\hskip 1em plus 0.5em
  minus 0.4em\relax Springer Science \& Business Media, 2007.

\bibitem{wing2015systemic}
J.~W. Wing, T.~N. Andrew, and D.~Petkov, ``A systemic framework for improving
  clients' understanding of software requirements,'' in \emph{ECIS 2015
  Proceedings at AIS Electronic Library (AISeL)}, 2015.

\bibitem{elsawah2015beyond}
S.~Elsawah, A.~McLucas, and M.~Ryan, ``Beyond why to what and how: the use of
  systems thinking to support problem formulation in systems engineering
  applications,'' in \emph{21st International Congress on Modelling and
  Simulation}, 2015.

\bibitem{ulrich2010critical}
W.~Ulrich and M.~Reynolds, ``Critical systems heuristics,'' in \emph{Systems
  approaches to managing change: A practical guide}.\hskip 1em plus 0.5em minus
  0.4em\relax Springer, 2010, pp. 243--292.

\bibitem{checkland_action_1998}
P.~Checkland and S.~Holwell, ``Action research: its nature and validity,''
  \emph{Systemic practice and action research}, vol.~11, no.~1, pp. 9--21,
  1998.

\bibitem{VolereTemplate}
J.~Robertson and S.~Robertson, ``Volere requirements specification template,''
  01 2000, available in:
  https://www.volere.org/templates/volere-requirements-specification-template/.

\bibitem{hayes2008unobtrusive}
T.~L. Hayes, F.~Abendroth, A.~Adami, M.~Pavel, T.~A. Zitzelberger, and J.~A.
  Kaye, ``Unobtrusive assessment of activity patterns associated with mild
  cognitive impairment,'' \emph{Alzheimer's \& Dementia}, vol.~4, no.~6, pp.
  395--405, 2008.

\bibitem{thew2018value}
S.~Thew and A.~Sutcliffe, ``Value-based requirements engineering: method and
  experience,'' \emph{Requirements Engineering}, vol.~23, no.~4, pp. 443--464,
  2018.

\bibitem{milne2012power}
A.~Milne and N.~Maiden, ``Power and politics in requirements engineering:
  embracing the dark side?'' \emph{Requirements Engineering}, vol.~17, no.~2,
  pp. 83--98, 2012.

\bibitem{dunne2013speculative}
A.~Dunne and F.~Raby, \emph{Speculative everything: design, fiction, and social
  dreaming}.\hskip 1em plus 0.5em minus 0.4em\relax MIT press, 2013.

\bibitem{temko2007acoustic}
A.~Temko, ``Acoustic event detection and classification,'' Ph.D. dissertation,
  Department of Signal Theory and Communications, Universitat Politecnica de
  Catalunya, Barcelona, Spain, 2007.

\bibitem{socoro2017anomalous}
J.~Socor{\'o}, F.~Al{\'\i}as, and R.~Alsina-Pag{\`e}s, ``An anomalous noise
  events detector for dynamic road traffic noise mapping in real-life urban and
  suburban environments,'' \emph{Sensors}, vol.~17, no.~10, p. 2323, 2017.

\end{thebibliography}

\end{document}